\documentclass[aps,twocolumn,nofootinbib,showpacs,pra,aps,10pt]{revtex4-2}
\usepackage[dvips]{graphicx}
\usepackage[english]{babel}
\usepackage[T1]{fontenc}
\usepackage{mathrsfs}
\usepackage[tbtags]{amsmath}
\usepackage{amssymb}
\usepackage{amsxtra}
\usepackage{xcolor}
\usepackage{amsopn}
\usepackage{latexsym}
\usepackage[mathcal]{eucal}
\usepackage{mathtools}

\newcommand{\BE}{\begin{equation}}
\newcommand{\EE}{\end{equation}}
\newcommand{\BA}{\begin{align}}
\newcommand{\EA}{\end{align}}

\newcommand{\nn}{\nonumber}

\begin{document}

\title{A reason why we do not observe Schr\"odinger's cats}

\author{Fabio Siringo}\email{fabio.siringo@ct.infn.it}

\affiliation{Dipartimento di Fisica e Astronomia 
dell'Universit\`a di Catania,\\ 
INFN Sezione di Catania,
Via S.Sofia 64, I-95123 Catania, Italy}

\date{\today}

\begin{abstract}
A reason is discussed (maybe not the only one) for why we do not see any superposition of macroscopic states in the real world.
Under the general assumption that quantum macrostates are statistical ensembles of microstates, it is shown that any superposition of macrostates is reduced in a very short time by the unitary dynamics of the ordinary Schr\"odinger equation, deducing
the Born rule without having to postulate it. 
In more detail, the macroscopic and microscopic degrees of freedom are decoupled in the Schr\"odinger
equation, yielding an effective stochastic equation for the macroscopic variables,
with the ensemble average of  the microscopic amplitudes that acts as a
self-generated internal white noise. The stochastic equation is shown to
be a reducing It\^o equation if some general causality conditions are met, predicting a very quick collapse of any macroscopic superposition upon formation,
with probabilities which satisfy the Born rule.
 In the context of the von Neumann measurement scheme, the relevance of the result is discussed as a simple dynamical solution of the measurement problem.
\end{abstract}

\maketitle
\newpage

\section{Introduction}
Despite the success of Quantum Mechanics (QM), with more than a century of experiments confirming its laws, the classical limit is not well understood yet. Decoherence, due to environment correlations\cite{Zurek}, does not fully explain why we don't observe superpositions of macroscopic objects; moreover, the Born rule must be {\it postulated} in order to predict the results of measurements. 

The collapse, which breaks the unitary time evolution, only occurs when a microscopic quantum system interacts with a macroscopic apparatus. From that point of view, the random reduction of any quantum superposition seems to be a special feature of
macroscopic objects, rather than an intrinsic aspect of particle dynamics. It is the macroscopic observer which fails to split into
different superposed macroscopic states. Then, the abrupt reduction of the microscopic wave functions would arise by their entanglement with the macroscopic observers: according to the von Neumann picture of the measurement process, the Born rule is satisfied if it holds for macroscopic objects. 

Focusing on the quantum behavior of macroscopic systems,
minor changes of the Schr\"odinger equation have been proposed by many authors, notably in Ref.\cite{Rimini0}, with the aim of solving the measurement problem by
a dynamical mechanism which would lead to a quick collapse of any superposition of macroscopic objects. A full account of the many attempts along that line can be found in Ref.\cite{Rimini}. More recently, a remarkable formal completion of QM\cite{baldo1,baldo2} has been proposed, which solves all the measurement problems at the cost of extending the numerical support of QM by nonstandard analysis. 

In this paper, we take a more modest and conservative point of view
and explore the hypothesis that the collapse of macroscopic superpositions might arise by the ordinary Schr\"odinger equation of the standard QM because of the
stochastic nature of any macrostate. That would give a reason (maybe not the only one) why we do not observe any Schr\"odinger's cat in the real world.

Actually, any dynamical reduction process must have to do with the stochastic microscopic nature of the macroscopic objects, since a {\it random}
collapse can only arise from the deterministic laws of QM if we describe the macroscopic system as a statistical ensemble, rather than as a single macroscopic quantum state. With plenty of disordered interactions around, it can be easily shown that the Schr\"odinger equation becomes a stochastic equation which might explain the randomness of the reduction process if some conditions are met.

In this paper it is shown  that, after having decoupled the macroscopic and microscopic degrees of freedom, the ensemble average of the microscopic variables acts as a self-generated internal white noise for the macroscopic amplitudes.
Then, the ordinary Schr\"odinger equation leads to a stochastic equation for the macroscopic degrees of freedom. The stochastic equation is shown to be a {\it reducing} equation, predicting a very quick collapse of the superposition with probabilities which satisfy the Born rule, if both the energy bandwidth $\Delta E$ and the space position width $\Delta X$, which are spanned by the microstates in a single macroscopic state, are large enough to comply with the constraint $\Delta E\, \Delta X\gg \hbar c$. Since that condition is fully satisfied by any ordinary macroscopic object, the Born rule would be {\it predicted} and would not need to be postulated in the measurement process, i.e. whenever a microscopic quantum system is entangled with a macroscopic experimental apparatus which would collapse almost instantly. But of course, some experimental evidence would be required in order to assess the validity of the assumptions made in the derivation.

 The paper is organized as follows: in Sec.II the microscopic and macroscopic degrees of freedom are decoupled for a generic superposition of macroscopic quantum states, deriving  an effective stochastic equation
 for the macroscopic variables; in Sec.III the random matrix element generated
 by the microscopic amplitudes is shown to act as a white noise; in Sec.IV the stochastic equation is studied as It\^o, the Fokker-Planck equation is derived
 for the probability distribution of the solutions and a reduction of the superposition is found, yielding the Born rule; in Sec.V the It\^o vs. Stratonovich interpretation of the stochastic equation is discussed, confirming that the It\^o assumption is correct if some conditions on the relaxation time of the system are met; In Sec.VI those conditions are shown to be fully satisfied by any real macroscopic system because of general causality
 arguments; in Sec.VII the relevance of the result is discussed as a simple dynamical solution of the measurement problem.

\section{The stochastic equation}

We can always write 
a generic superposition of macroscopic quantum states  as
\begin{equation}
|\psi\rangle=\sum_n\sum_\alpha b_n^\alpha\, |\phi_n^\alpha\rangle
\label{superp}
\end{equation}
where the orthonormal complete set $\{\, |\phi_n^\alpha\rangle\, \}$ is a set of macroscopic many-particle states for the whole $N$-particle system. The lower index, $n$, refers to the macroscopic properties of the system while the upper index, $\alpha$,
refers to the specific quantum microstate.
States with a different 
index $n$ differ for some macroscopic observable which can be measured, like total energy $E_n$, center of mass position ${\bf R}_n$, total momentum ${\bf P}_n$, etc. The states are not exact eigenstates of the observables since, in the classical limit, we find states that
are {\it almost} eigenstates of all the  observables. We can divide the eigenvalue spectra of the observables in very small discrete bins and assume that in the state $|\phi_n^\alpha\rangle$ the generic observable $\hat A$ has an expectation value which falls in the bin $\Delta A_n$ and  takes a value $A\approx A_n$.
By definition, two macrostates are perceived as different
if at least one of the observables falls in different bins.
On the other hand, there is a huge number of different microscopic realizations of the same macrostate, labeled by the Greek upper index: two states $|\phi_n^\alpha\rangle$ and $|\phi_n^\beta\rangle$ are many-particle states which differ by microscopic details
but are perceived as identical.

Let us define the macroscopic squared amplitudes
\begin{equation}
p_n=\sum_\alpha |b_n^\alpha|^2, \quad \sum_n \,p_n=1\, 
\label{xn}
\end{equation}
and the complex microscopic amplitudes 
\begin{equation}
\eta_n^\alpha=\frac{b_n^\alpha}{\sqrt{p_n}}
\end{equation}
which satisfy
\begin{equation}
\sum_\alpha |\eta_n^\alpha|^2=\sum_\alpha{\frac{|b_n^\alpha|^2}{p_n}}=1\, , \qquad 0\le |\eta_n^\alpha|\le 1
\label{eta}
\end{equation}
and inherit the phases
of the coefficients $b_n^\alpha$.
In the language of the Born rule, $p_n$ would be the probability of collapsing in the single macrostate
$|\psi_n\rangle$
\begin{equation}
|\psi\rangle=\sum_n \sqrt{p_n}\, |\psi_n\rangle\quad\longrightarrow\quad
|\psi_n\rangle=\sum_\alpha \eta_n^\alpha\, |\phi_n^\alpha\rangle\, .
\label{superp2}
\end{equation}
There is a huge number $N_n$ of equivalent microstates in each macrostate $n$, then the  set $\{\eta_n^\alpha\}$
must be regarded as a statistical ensemble of stochastic
complex variables with the only  constraint of Eq.(\ref{eta}) and configurational averages
\begin{equation}
    \langle \eta_n^\alpha\rangle=0\,,\quad  \langle\, \eta_n^\alpha\,{\eta_m^\beta}^*\, \rangle=\delta_{nm}\delta_{\alpha\beta}\,\frac{1}{N_n}\, .
\label{etaave}
\end{equation}
We do not know
the coefficients $\eta_n^\alpha$ and we cannot measure them any way. According to
decoherence\cite{Zurek} we don't keep track of the random phases in the macroscopic amplitudes $\sqrt{p_n}$ which are taken as real in Eq.(\ref{superp2}).
Thus, decoherence emerges naturally from the randomness of the microscopic 
amplitudes $\eta_n^\alpha$, even without including an explicit interaction with the environment.

The standard Schr\"odinger equation
\begin{equation}
i\hbar \,\frac{\partial}{\partial t}|\psi(t)\rangle=\hat H\,|\psi(t)\rangle
\label{Schr1}
\end{equation}
can be written as a dynamical equation for the amplitudes $b_n^\alpha (t)$
\begin{equation}
i\hbar \,\frac{d b_n^\alpha}{dt}=\sum_m\sum_\beta    \langle \phi^\alpha_n|\hat H\,|\phi_m^\beta\rangle \, b_m^\beta(t)\, .
\label{Schr0}
\end{equation}
Let us split the Hamiltonian $\hat H$ in a block diagonal part $\hat H_0$ and a pure off-diagonal interaction 
$\hat V=\hat H-\hat H_0$ with matrix elements
\begin{equation}
 {H_0\,}^{\alpha\beta}_{nm}=\delta_{nm}\,\langle \phi^\alpha_n|\hat H\,|\phi_n^\beta\rangle
\,, \
V_{nm}^{\alpha\beta}=(1-\delta_{nm})\,
\langle \phi^\alpha_n|\hat H\,|\phi_m^\beta\rangle\, .\quad
\nn
\end{equation}
In the interaction picture, the Schr\"odinger equation, Eq.(\ref{Schr0}), can be written as 
\begin{equation}
\frac{d}{dt}|b_n^\alpha(t)|^2=\frac{1}{i\hbar}
\sum_m \sum_\beta\left[
{b_n^\alpha}^* \,b_m^\beta\,V_{nm}^{\alpha\beta}- \,c.c.\right]\, .
\label{stoch}
\end{equation}
and can be summed over the Greek index $\alpha$ of the coefficients $b^\alpha_n$, yielding a dynamical equation for the macroscopic amplitudes $p_n(t)$
\begin{equation}
\frac{d p_n}{dt} =
\frac{1}{i\hbar}\sum_m \sqrt{p_n\,p_m}\, \Big[W_{n\,m}-W_{m\, n}\Big]
\label{stoch1}
\end{equation}
where $W_{n\,m}$ is the complex Hermitian matrix element
\begin{equation}
W_{n\,m}(t)=
\sum_{\alpha\beta} \,{\eta_n^\alpha}^*\, 
V_{nm}^{\alpha\beta}(t)\
\eta_m^\beta\, ,
\label{Wt}
\end{equation}
and $V_{nm}^{\alpha\beta}(t)$ is the time-dependent matrix element
of $\hat V(t)=e^{i\hat H_0\, t/\hbar}\,\hat V\,
e^{-i\hat H_0\, t/\hbar}$.
While Eq.(\ref{stoch1}) is an exact dynamical equation, we are going to show right away that the random matrix element
$W_{n\, m}$ acts as a self-generated Hermitian white noise whose ensemble averages
satisfy
\begin{align}
&\langle W_{nm}\rangle=0\, ,\nn\\ 
&\langle W_{nm}(t)W_{m'n'}(t')\rangle=\delta_{nn'}\, \delta_{mm'}\,d_{nm}(t-t')\, ,\nn\\
&d_{nm}(t-t')\approx \sigma_{nm}^2\,\delta(t-t')\, .
\label{whitenoise}
\end{align}
Thus, Eq.(\ref{stoch1}) must be studied as a stochastic differential equation
(SDE) for the macroscopic variables $p_n$. The equation belongs to the class of reducing SDE discussed in Ref.\cite{Pearle79}: if studied as a It\^o SDE the equation predicts the random collapse of the superposition to one of the single superposed states $|\psi_n\rangle$, with a probability given by the initial value $p_n(0)$, in agreement with the Born rule.

While Eq.(\ref{stoch1}) is exact and arises from the ordinary Schr\"odinger equation for the whole quantum macrostate, it cannot be 
easily derived for a single microscopic quantum system, even in presence of an external noise, unless some non-linear term were added by
hand to the ordinary dynamical equation\cite{Pearle79,Pearle86}.
In fact, when written as a SDE for an effective amplitude $b_n=\sqrt{p_n}$, Eq.(\ref{stoch1}) is equivalent to the It\^o SDE
\begin{equation}
{d b_n}=\frac{1}{2i\hbar}\sum_m(W_{n\,m}-W_{m\, n})\, b_m\, dt 
-\sum_m \frac{\sigma_{nm}^2}{4\hbar^2}\frac{b_m^2}{b_n}\, dt\, .
\label{stochequiv}
\end{equation}
The non-linear last term must be added because of the It\^o lemma: in 
going back to Eq.(\ref{stoch1}), the stochastic differential $dp_n=d(b_n^2)$
would acquire a second order It\^o
correction term,
\begin{equation}
d(b_n^2)=2 \,b_n\,db_n +db_n\,d b_n\, 
\label{db2}
\end{equation}
where, because of the white noise correlators, Eq.(\ref{whitenoise}),
\begin{equation}
db_n\, db_n= \sum_m \frac{\sigma_{nm}^2}{2\hbar^2}\,{b_m^2}\ dt
\end{equation}
and is first order in time. The non-stochastic drift term which is added
in Eq.(\ref{stochequiv}) cancels the It\^o correction in Eq.(\ref{db2})
thus recovering Eq.(\ref{stoch1}). Without the non-linear term, the
Schr\"odinger-like Eq.(\ref{stochequiv}) would lead to a different
non-reducing equation for $p_n$ and would not predict any collapse.
Thus, it is remarkable that, for a superposition of macrostates, 
Eq.(\ref{stoch1}) arises from the exact dynamical equation without having to add any non-linear term by hand.

\section{White Noise}

Since we would like to study Eq.(\ref{stoch1}) as a It\^o SDE, we must first show that the self-generated random matrix element $W_{n\, m}$ can be approximated by an Hermitian white noise which 
satisfies Eq.(\ref{whitenoise}).

In order to make explicit the time dependence of the matrix element 
$V_{nm}^{\alpha\beta}(t)$ in Eq.(\ref{Wt}), let us introduce a complete
set of eigenstates for $\hat H_0$.
Being block diagonal, $\hat H_0$ must have a different subset of eigenstates $\{\, |n,k\rangle\,\}$ in each bin $n$,
satisfying 
\begin{equation}
\hat H_0\,|n,k\rangle=E_k^{(n)}\, |n,k\rangle\, , \qquad
\sum_k |n,k\rangle\langle n,k|={\cal P}_n
\end{equation}
where ${\cal P}_n$ is a projector onto the subspace spanned by $|\phi_n^\alpha\rangle$. Denoting
$\gamma_{n,k}^\alpha=\langle n,k|\phi_n^\alpha\rangle$, 
we can write in each bin
\begin{equation}
\sum_\alpha {\gamma_{n,k}^\alpha}^*\, \gamma_{n, k'}^\alpha 
= \sum_\alpha\langle n,k'|\phi_n^\alpha\rangle \langle \phi_n^\alpha|n,k\rangle =\delta_{k,k'}\, .
\label{gg}
\end{equation}
Inserting  the matrix element
\begin{equation}
V_{nk,\,mk'}(t)=\langle n,k|\,\,e^{i\hat H_0\, t/\hbar}\,\hat V\,
e^{-i\hat H_0\, t/\hbar}\,\, |m,k'\rangle\, , 
\end{equation}
the matrix $W_{n\,m}$ reads
\begin{widetext}
\begin{equation}
W_{n\,m}(t)=
\sum_{\alpha\beta}{\eta_n^\alpha}^*\,\eta_m^\beta\,\sum_{k,k'}
\left[{\gamma_{n,k}^\alpha}{\!^*}\ V_{nk,\,mk'}(0) \ \gamma_{m,k'}^\beta\right]
\ e^{i(E^{(n)}_k-E^{(m)}_{k'})\, t/\hbar}\, .
\end{equation}
In the interaction picture, we expect that the strongest time dependence arises from the wildly oscillating phase factors due to $\hat H_0$ and that we can neglect the weak time dependence of the amplitudes for very short time intervals. In other words, assuming that the variables $\eta_n^\alpha(t)$ are still correlated  after a very brief time interval $t'-t\approx \tau_c$, we can set  
$\langle\,{\eta_n^\alpha}^*(t)\, \eta_m^\beta(t')\rangle\approx
\langle\,{\eta_n^\alpha}^*(t)\, \eta_m^\beta(t)\rangle=
\delta_{nm}\delta_{\alpha\beta}\ \frac{1}{N_n}$.  Since we are going to find correlation times of order $10^{-22}~$s, shorter than the electron Compton time $\hbar/2mc^2$, the approximation is hardly a limit for any macroscopic object. Using Eq.(\ref{gg}),
the correlator then reads 
\begin{equation}
\langle\, W_{n\, m}(t)\, W_{m'\, n'}(t') \,\rangle=
\frac{\delta_{nn'}\,\delta_{mm'}}{N_nN_m}\sum_{kk'}
|V_{nk,\,mk'}|^2 
e^{i(E^{(n)}_k-E^{(m)}_{k'})\, (t-t')/\hbar}\, .
\end{equation}
The sum is over a huge set of $N_n\times N_m$ energy differences, with $E^{(n)}_k$ ranging in a limited bin domain $\Delta E$ around the average value $E_n$. We can write the sum as an integral over the energy difference
${\cal E}=(E^{(n)}_k-E_n)-(E^{(m)}_{k'}-E_m)$ 
\begin{equation}
\langle\, W_{n\, m}(t)\, W_{m'\, n'}(t') \,\rangle
\approx
\delta_{nn'}\,\delta_{mm'}\, e^{i(E_n-E_m)(t-t')/\hbar}\, 
\left[\frac{1}{2\Delta E} \int_{-\Delta E}^{\Delta E} d{\cal E}\,\Big|V_{nm}({\cal E})\Big|^2\, 
e^{i{\cal E}(t-t')/\hbar}\right]
\end{equation}
\end{widetext}
where $\big|V_{nm}({\cal E})\big|^2$ is the average squared matrix element for a given energy difference ${\cal E}$. Defining $\tau_c=\hbar/\Delta E$, on general grounds, the integral gives a peaked correlator $d_{nm}(t-t')$ with a width of order $\tau_c$. Since $\tau_c$ is very small for a macroscopic $\Delta E$, Eq.(\ref{whitenoise}) is satisfied 
with
\begin{equation}
\sigma_{nm}^2\approx\frac{\pi\hbar\, |V_{nm}(0)|^2}{\Delta E}.
\end{equation}
We observe that the bin size $\Delta E$ must be large enough to grant that we can perceive as different the two energies $E_n$ and $E_n+\Delta E$; say much larger
than the quantum uncertainty $\sigma_E$ which, for an assembly of $N$ 
microscopic elements can be estimated to be larger than the Poissonian 
$\sigma_E/E\approx 1/\sqrt{N}$. Assuming a macroscopic number $N\approx N_A=0.6\cdot 10^{24}$ and an energy of order unity, then $\Delta E\gg 10^{-12}~\text{J}~\approx 1~$MeV, yielding a very tiny correlation time
$\tau_c=\hbar/\Delta E\ll \hbar/2mc^2=\tau_0$ where the electron Compton time, $\tau_0=10^{-22}$~s, is the shorter time that we may consider for the dynamics of electrons.
Since matter is basically made of electrons, the limit $\tau_c\to 0$ can be safely taken and the matrix $W_{n\, m}$ can be approximated by a pure
white noise with ensemble averages given by Eq.(\ref{whitenoise}).

\section{Stochastic reduction}

For a very small but yet perceivable bin width $\Delta E$, the correlation time $\tau_c$ is many orders of magnitude shorter than the relaxation time $\tau_R$ of the macroscopic variables $p_n$, which cannot
follow the fast dynamics of the microscopic
amplitudes. As shown in Ref.\cite{k}, in the present regime, $\tau_R\gg\tau_c$, the stochastic equation,
Eq.(\ref{stoch1}), must be studied as a It\^o SDE rather than a Stratonovich one
as is usually assumed instead for a microscopic quantum system with a colored external noise. In the next section we will address the issue of the It\^o vs. Stratonovich interpretation in more detail. 

Assuming for a while that Eq.(\ref{stoch1}) can be studied as a It\^o SDE, here we show  that the stochastic equation belongs to a larger well-known class of reducing equations.
When regarded as a It\^o SDE, Eq.(\ref{stoch1}) can be studied by the method of Ref.\cite{Pearle79}: the ensemble of solutions must satisfy  
the Fokker-Planck diffusion equation 
\begin{equation}
\frac{\partial \rho}{\partial t}=-\sum_i\frac{\partial}{\partial p_i} \left[\frac{\langle dp_i\rangle}{dt} \rho\right]
+\frac{1}{2}\sum_{ij}\frac{\partial^2}{\partial p_i\partial p_j} \left[\frac{\langle dp_id p_j\rangle}{dt} \rho\right]
\label{FP}
\end{equation}
where $\rho\equiv\rho\big(p_n(t),t; p_n(0)\big)$ is the probability density 
of the solution set $\{\, p_n(t)\,\}$ with initial values $\{\,p_n(0)\,\}$. 
The first and second moments are easily evaluated by Eq.(\ref{stoch1}) which,
if regarded as It\^o, gives no drift term, $\langle dp_i\rangle=0$, and,
by inserting Eq.(\ref{whitenoise}) in the averages
\begin{align}
 &\langle dp_i dp_j\rangle=\nn\\
 & -\frac{dt}{\hbar^2} \sum_{mk}
 \sqrt{p_i\, p_j\,p_m\,p_k}\,\left[ 2\,\delta_{ik}\,\delta_{jm}\,\sigma_{im}^2
-2\,\delta_{ij}\,\delta_{mk}\,\sigma_{im}^2\right]\nonumber\\
&= -\frac{2\,dt}{\hbar^2} \sum_{m}\sigma_{im}^2\,
{p_i\, p_m}\,\left(\delta_{jm}-\delta_{ij}\right)\, .
\end{align}
Substituting in Eq.(\ref{FP}), the Fokker-Planck equation reads
\begin{align}
\frac{\partial \rho}{\partial t}&=-\frac{1}{\hbar^2}
\sum_{ijm}\sigma_{im}^2\left(\delta_{jm}-\delta_{ij}\right)
\frac{\partial^2}{\partial p_i\partial p_j} 
\big(p_i\,p_m\,\rho\big)\nn\\
&=\frac{1}{\hbar^2}
\sum_{ij}\sigma_{ij}^2\left(
\frac{\partial^2}{\partial p_i^2}-
\frac{\partial^2}{\partial p_i\partial p_j} 
\right)
\big(p_i\,p_j\,\rho\big)
\label{FP2}
\end{align}
and using the exchange symmetry of $\sigma_{ij}^2\,p_i\,p_j$
\begin{equation}
  \frac{\partial \rho}{\partial t}=\frac{1}{2\hbar^2}
\sum_{ij}\sigma_{ij}^2\left(
\frac{\partial}{\partial p_i}-\frac{\partial}{\partial p_j}
\right)^2
\big(p_i\,p_j\,\rho\big)\,. 
\label{FP3}  
\end{equation}
This is exactly the same diffusion equation found in Ref.\cite{Pearle79} upon integration over the phases. The equation is known to be a {\it reducing} equation: multiplying by $(p_k p_m)$, with $k\not=m$, and integrating over all the stochastic
variables, with a volume element $d\Gamma=\prod_i dp_i$, observing that
\begin{equation}
\frac{\partial }{\partial t}\int \rho\, p_k\,p_m\, d\Gamma=
\frac{\partial }{\partial t}\langle p_k\, p_m\rangle\, ,
\end{equation}
we find from Eq.(\ref{FP3})
\begin{align}
&\frac{\partial }{\partial t}\langle p_k\, p_m\rangle=\nn\\
&\frac{1}{2\hbar^2}
\sum_{ij}\sigma_{ij}^2 \int p_k\, p_m
\left(
\frac{\partial}{\partial p_i}-\frac{\partial}{\partial p_j}
\right)^2
\big(p_i\,p_j\,\rho\big)\,d\Gamma\,, 
\label{pp}  
\end{align}
then, integrating by parts twice
\begin{equation}
\frac{\partial }{\partial t}\langle p_k\, p_m\rangle=
-\frac{2\sigma_{mk}^2}{\hbar^2}\int p_k\, p_m\, \rho\, d\Gamma=
-\frac{2\sigma_{mk}^2}{\hbar^2}\langle p_k\, p_m\rangle\, .
\end{equation}
This equation has the exponential decreasing solution for $k\not=m$
\begin{equation}
\langle p_k(t)\,p_m(t)\rangle=\langle p_k(0)\,p_m(0)\rangle\, e^{-2\sigma_{km}^2\, t\,/\,\hbar^2}
\label{exp}
\end{equation}
which leads to a collapse of the superposition in the very short time scale 
\begin{equation}
t\approx \hbar^2/\sigma_{km}^2\approx\frac{\hbar\Delta E}{|V_{km}(0)|^2}\,.
\label{collapse}
\end{equation}
In fact, it is evident that if $\langle p_k\, p_m\rangle\to 0$ for all pairs $k\not=m$, then only one randomly chosen $p_m(t)$ can differ from zero after that short reducing time: all $p_k(t)$ must vanish exponentially, except for a randomly chosen one which tends to 1, preserving the normalization sum $\sum_n p_n=1$, as expected from the unitarity of the time evolution.
As observed in Refs.\cite{Pearle79,Pearle86},
on average, the observed probability $P_m$ of finding the system in the state $|\psi_m\rangle$ can be expressed as the final value of 
\begin{equation}
\langle p_m(t)\rangle=1\cdot P_m+0\cdot (1-P_m)=P_m\, .
\end{equation} 
Then, the Born rule is recovered because, according to Eq.(\ref{whitenoise}),
in the It\^o interpretation, the average of Eq.(\ref{stoch}) gives $d\langle p_m(t)\rangle/dt=0$ (there is no drift term) and the probability $P_m=\langle p_m(t)\rangle$ is equal to the initial value $p_m(0)$, as predicted by the Born rule.

\section{It\^o vs. Stratonovich}

It is widely believed that the random noise generated by  real macroscopic systems and by the environment must be regarded
as a {\it colored} noise, with the delta function $\delta(t-t')$ in Eq.(\ref{whitenoise}) replaced by a peaked correlator $d_{nm}(t-t')$, with a finite width $\tau_c\sim \hbar/\Delta E$ where $\Delta E$ is the energy
bandwidth of the stochastic process. 
The finite correlation time $\tau_c$ might have a disastrous effect on the
stochastic equation: a drift term should be added to the right-hand side of Eq.(\ref{stoch1}), leading to an average $d\langle p_n\rangle /dt\not=0$, spoiling  the reduction process entirely.
Technically, we say that the stochastic equation is studied as a Stratonovich SDE\cite{strato} rather than a Itô equation\cite{ito} and, as such, it does not lead to  any collapse of the initial state $| \psi\rangle$,
in contrast to the experimental evidence.

In presence of multiplicative noise, the correct interpretation of a 
stochastic equation, the It\^o vs. Stratonovich debate\cite{IvS1,IvS2,IvS3,IvS4}, has been discussed for several years. The noise is {\it multiplicative} in Eq.(\ref{stoch1})
because the white noise $W_{nm}$ has a factor $\sqrt{p_np_m}$ which depends on the stochastic variables $p_n(t)$. As shown in Ref.\cite{k}, any real process of that kind is characterized by two different time scales: the correlation time $\tau_c$ of the colored noise and the inertial relaxation time of the system $\tau_R$. While both times are usually sent to zero
as a reasonable approximation, the order of the two limits cannot be interchanged. There is  analytical and numerical evidence by now\cite{k} that the resulting stochastic equation must be regarded as a Stratonovich SDE when $\tau_c\gg\tau_R$
and as a It\^o SDE when $\tau_c\ll\tau_R$.

In the first case, the occurrence of an extra drift term, which spoils the reduction process, can be easily understood by the following perturbative argument. For semplicity, let us consider a diagonal random noise $W_{n\, m} =\delta_{nm}\, W$ and the decoupled Schr\"odinger-like equation
\begin{equation}
i\hbar\, \frac{d b}{dt}=W\, b  \,.
\label{SchrLike}
\end{equation}
If we integrate and iterate, we can formally
expand the average 
$\langle \Delta b(t)\rangle=\langle b(t)-b(0)\rangle$
as
\begin{align}
&\langle \Delta b(t)\rangle=\frac{1}{ i\hbar}  b(0) \int_0^t\langle W(t')\rangle\,dt'\nn\\
&+\frac{1}{(i\hbar)^2} b(0) \int_0^tdt'\int_0^{t'} dt''\,\langle W(t') \,W(t'')\rangle\nn\\
&+{\cal O}(\langle W\cdot W\cdot W\rangle)
\label{expansion}
\end{align}
with strict time ordering, $t'\ge t''$. 
Using Eq.(\ref{whitenoise}), the first order term vanishes while the second order term picks up a contribution 
around $t'\approx t''$ where the integrals overlap if the width $\tau_c$ of the correlator
\begin{equation}
\langle W(t) \,W(t')\rangle\approx \sigma^2\,\delta(t-t')  
\end{equation}
is finite. The double integral can be approximated as $\sigma^2\,t\,/2$, half of the contribution of the white noise delta distribution at the upper extreme $t'$, yielding a finite drift
velocity $\langle \Delta b\rangle /t\not=0$ in the limit $t\to 0$
\begin{equation}
\frac{d}{dt}\langle \Delta b(t)\rangle=
\frac{\langle \Delta b(t)\rangle}{t}=
-\frac{\sigma^2 \,b(0)}{ 2\hbar^2} \,.
\label{dbdt1}
\end{equation}
The argument is valid if $\tau_R=0$ and the variables
$b(t)$ and $db/dt$ are strictly evaluated at the same time in Eq.(\ref{SchrLike}), otherwise, if a delay time $\tau_R>\tau_c$ were considered, the integrals in the second order term would not overlap.
This second case can be addressed by adding a small inertial term to Eq.(\ref{SchrLike}), as discussed in Ref.\cite{k}, writing
the more realistic stochastic equation 
\begin{equation}
\tau_R\, \ddot b=i\,  \dot b-\frac{1}{\hbar}W\,b
\label{inertial}
\end{equation}
where the dots represent derivatives with respect to time.
This modified equation
resembles the classical equation of motion for a damped particle with inertial mass equal to $\tau_R$.
The Schr\"odinger equation, Eq.(\ref{SchrLike}), can be seen as
the approximate version of the above equation in the limit $\tau_R\to 0$. In that limit, the ``velocity'' $\dot b$
is instantly proportional to the ``force'' as it occurs in
a stationary classical regime. When the inertial ``mass'' $\tau_R$ is not zero, it is the acceleration $\ddot b$ which
is proportional to the force, while the velocity is delayed
by the inertial term. As shown in Ref.\cite{k}, if the limit
$\tau_c\to 0$ is taken first, while keeping $\tau_R\not= 0$, the resulting stochastic process is well described by the corresponding It\^o SDE, as expected by the expansion 
(\ref{expansion}) when $\tau_R\gg \tau_c$. Even when $\tau_R$
is eventually sent to zero, recovering the Schr\"odinger equation, no extra drift term must be added if the limit is taken while keeping
$\tau_R\gg\tau_c$, since the amplitude
$b$ does not follow the dynamics of the noise instantly. 

While the formal proof in Ref.\cite{k} was derived for a real dissipative term, we show that the same argument holds in presence of the imaginary coefficient of 
$\dot b$ in the Schr\"odinger equation.
Using the method of Ref.\cite{k}, by the variation of constants formula, we solve for $\dot b$
\begin{equation}
\dot b(t)=\dot b(0)\, e^{it/\tau_R}\, -\frac{1}{\hbar\tau_R}\int_0^t dt'\, e^{i(t-t')/\tau_R}\ W(t')\,b(t')\, .    
\label{dotb}
\end{equation}
Integrating further, by parts, the displacement $\Delta b(t)=b(t)-b(0)$ reads
\begin{equation}
\Delta b(t)= \frac{1}{i \hbar}\int_0^t dt'\, 
W(t')\,b(t')\,  +\, {\cal O}(\tau_R)\, ,     
\label{db}
\end{equation}
where the ${\cal O}(\tau_R)$ terms include the integral
\begin{equation}
-\frac{1}{i \hbar}\int_0^t dt'\, e^{i(t-t')/\tau_R}\ W(t')\,b(t') \to 0 \, ,    
\label{lebesgue}
\end{equation}
which vanishes in the limit $\tau_R\to 0$ because of the Riemann-Lebesgue lemma if
$W(t)$ is a smooth colored noise. For a white noise, with $\tau_c=0$, the integral
vanishes anyway by the same path of the integral in Eq.(\ref{db}), as shown below.
Actually, because of the essential singularity in $\tau_R=0$, the limit must be handled with care; here, Eq.(\ref{db}) leads to the correct limit as shown by a differentiation
that gives back the Schr\"odinger equation,
$i\hbar\, \dot b=W\,b$, which is the limit of Eq.(\ref{inertial}) for $\tau_R\to0$.

By the same argument of Eq.(\ref{expansion}), we can take the configurational average
of Eq.(\ref{db}) and iterate, inserting $b(t)=b(0)+\int_0^t\dot b(t')\, dt'$, with
$\dot b$ given by Eq.(\ref{dotb}). Neglecting terms of order ${\cal O}(\tau_R)$ and higher powers of $t$
\begin{align}
&\langle\Delta b(t)\rangle=
-\frac{b(0)}{i \hbar^2\tau_R}\times\nn\\
&\int_0^t dt' \int_0^{t'} dt''\int_0^{t''} dt'''  
e^{i(t''-t''')/\tau_R}
\langle W(t')W(t''')\rangle\, .    
\label{dbav}
\end{align}
At variance with Eq.(\ref{expansion}), there is an internal integration on $t''$
which prevents the overlap for a pure white noise.

The integrals can be evaluated exactly if we approximate the colored noise
by an exponential correlator
\begin{equation}
\langle W(t)W(t')\rangle=\sigma^2\,\frac{\lambda}{2}\,e^{-\lambda|t-t'|}\to \sigma^2\,\delta(t-t')\, 
\label{colorednoise}  
\end{equation}
where the limit $\lambda=1/\tau_c\to \infty$ will be taken at the end of the calculation. Taking a time derivative, the result is
\begin{equation}
\frac{d}{dt}\langle \Delta b(t)\rangle=
-\,\frac{\sigma^2 \,b(0)}{ 2\hbar^2} \
\frac{A(t)}{1+i\lambda\tau_R}\, , \quad A(t)\to 1
\label{dbdt}
\end{equation}
where the time dependent factor $A(t)$ is
\begin{equation}
A(t)=\left\{1-e^{-\lambda t}\left[1
-i\lambda\tau_R\left( e^{it/\tau_R}-1\right) \right]
\right\}
\end{equation}
and tends to 1 in the double limit $\tau_R\to 0$, 
$\tau_c=1/\lambda\to 0$, irrespective of the limit order. 
As expected, the final result in Eq.(\ref{dbdt}) depends on the product $\lambda\tau_R=\tau_R/\tau_c$
and the limits cannot be interchanged.
If $\tau_R\ll\tau_c$ we can neglect $\lambda\tau_R\ll 1$
and Eq.(\ref{dbdt}) gives the same identical drift
term of Eq.(\ref{dbdt1}); if $\tau_R\gg \tau_c$, we can safely take
the white noise limit $\lambda\,\tau_R\to \infty$ and the drift term
goes to zero. Thus, the stochastic equation must be studied as a It\^o
SDE, without any added drift term, when the correlation time of the
colored noise is much shorter than the inertial relaxation time.

In the stochastic equation for the macroscopic variables $p_n$, 
Eq.(\ref{stoch1}), the double limit $\tau_c\to 0$ and $\tau_R\to 0$ is
understood, as a very good approximation to the real dynamics of the system.
Thus, in order to establish the nature of the stochastic equation, It\^o vs. Stratonovich, we must restore the finite tiny times $\tau_R$, $\tau_c$, and 
compare them. While a correlation time $\tau_c\ll 10^{-22}$~s has been estimated in 
Sec. III for a macroscopic system, the correct inertial time $\tau_R$ can only be extracted by the dynamics. However, as we are going to discuss, causality suggests that $\tau_R\gg\tau_c$  for the macroscopic variables $p_n$.

\section{Causality}

We have seen that the reducing or non-reducing character of the
stochastic equation, Eq.(\ref{stoch1}),
depends on the existence of a small inertial relaxation term $\tau_R$ in the
dynamical equation and on its comparison with the correlation time of the actual colored noise.
It might seem that $\tau_R=0$, exactly, in Eq.(\ref{stoch1})
because no second derivative or inertial term appears explicitly in the Schr\"odinger equation.  However, in its general form of Eq.(\ref{Schr1}), the 
exact Schr\"odinger equation
does not say anything about the dynamics and the relaxation time of the system, unless the content of the operator $\hat H$ is specified. Eq.(\ref{Schr1}) just says that $\hat H$ is the generator of time evolution. Moreover, Lorentz covariance
is hidden in the Hamiltonian formalism and the principle of relativistic causality is violated in the non-relativistic context.

In a real many body system, because of relativistic causality, any change $\Delta X$ of the macroscopic linear size would require a time of order $\Delta X/c$ at least, as discussed later in  more detail, below Eq.(55).
Then, for $\Delta X\sim 1$~nm, we expect a relaxation time $\tau_R\sim 10^{-17}$~s, which is a negligible time but still many orders of magnitude larger than the estimated correlation time $\tau_c\ll10^{-22}$~s. Moreover, many processes  might generate inertial effects\cite{lieb0, lieb1,lieb2} with effective light cone 
causality driven by much smaller velocities than the speed of light. In general,
since matter is basically made of electrons, we argue that no macroscopic system can have a relaxation time smaller than the electron Compton time 
$\tau_0=\hbar/2mc^2\approx 10^{-22}$~s.

The real relaxation time of a macroscopic system must be extracted by the actual dynamics and 
does not appear explicitly in the Schr\"odinger equation, Eq.(\ref{Schr1}), unless
the content of the Hamiltonian is specified.
For instance, even at the level of a single particle,
the  relativistic Salpeter equation 
\begin{equation}
i\hbar \,\frac{\partial}{\partial t}|\psi(t)\rangle=
\left[\sqrt{\hat {\bf p}^2c^2+m^2c^4}+\hat V\right]|\psi(t)\rangle
\label{Salp}
\end{equation}
leads to causality violations at time scales
shorter than the Compton time $\tau_0$\cite{salpeter}. The problem is solved by inclusion of the negative energy solutions that arise by the  second order time derivative of the Klein-Gordon equation. In fact, a tiny inertial term appears in the non-relativistic limit if we square the Salpeter equation and recast it as a Klein-Gordon equation\cite{KG}
\begin{equation}
\left[i\hbar \,\frac{\partial}{\partial t}-\hat V\right]^2|\psi(t)\rangle=
\left[\hat {\bf p}^2c^2+m^2c^4\right]|\psi(t)\rangle\, .
\label{KG}
\end{equation}
If we substitute $|\psi(t)\rangle=|\phi(t)\rangle\, e^{-imc^2\, t/\hbar}$
and evaluate the square,
neglecting all terms of order $\hat V/mc^2$,  we obtain a modified
Schr\"odinger equation
\begin{equation}
i\hbar \,\frac{\partial}{\partial t}|\phi(t)\rangle=
\left[\frac{\hat {\bf p}^2}{2m}+\hat V\right]\,|\phi(t)\rangle
+\frac{\hbar^2}{2mc^2}\frac{\partial^2}{\partial t^2} |\phi(t)\rangle
\label{nr}
\end{equation}
which is still {\it exact} in the limit $\hat V\to 0$. The standard
non-relativistic equation is obtained if we can neglect the Compton time 
$\tau_0=\hbar/{2mc^2}\to 0$.
In the basis of the eigenstates $\{\,|k\rangle\,\}$ of $\hat {\bf p}$,
expanding $|\phi(t)\rangle=\sum_k b_k(t)\, |k\rangle$, the equation reads
\begin{equation}
\tau_0 \,\ddot b_k=i\dot b_k-\Omega_k\, b_k-\frac{1}{\hbar}\sum_q V_{k\,q}\, b_q
\label{nrexpl}
\end{equation}
where $\Omega_k=\hbar k^2/{2m}$ and $V_{k\,q}=\langle k|\hat V|q\rangle$.
We find the same structure of the modified dynamical equation introduced in Eq.(\ref{inertial}),
with a relaxation time $\tau_R=\tau_0$ which arises as a relativistic effect.
For electrons, $\tau_0\approx 10^{-22}$s and can be safely neglected in the non-relativistic context. However, the role of $\tau_0$ is crucial for the relativistic causality: in the
limit $\hat V\to 0$, Eq.(\ref{nrexpl}) has solutions 
$ b_k(t)=b_k(0)\, e^{-i\omega t}$ with the exact spectrum (with positive and negative energies)
\begin{equation}
\hbar\omega=-\frac{\hbar}{2\tau_0}\pm\hbar\sqrt{\frac{\Omega_k}{\tau_0}+\frac{1}{4\tau_0^2}}= -mc^2\pm\sqrt{m^2c^4+(\hbar k)^2 c^2}  
\end{equation}
while, if we just set $\tau_0=0$ in Eq.(\ref{nrexpl}), we would get the non-relativistic spectrum $\hbar\omega=-mc^2+\hbar^2k^2/2m$ and no light cone limit for the speed of particles. Thus, the inertial term in Eq.(\ref{nrexpl}) has important effects even at energies which are smaller than the mass gap $2mc^2$. 

For an extended macroscopic system, the relativistic causality requires even more stringent general constraints. 
The quantum expectation value of the position $\hat X$ for the superposition $|\psi\rangle$ can be expressed
as
\begin{equation}
\langle \psi |\hat X|\psi\rangle=\sum_{n\, m} \sqrt{p_np_m}\sum_{\alpha\beta} {\eta_n^\alpha}^\star\eta_m^\beta
\langle\phi_n^\alpha|\hat X|\phi_m^\beta\rangle 
\end{equation}
and, using Eq.(\ref{etaave}), it would be perceived as the time dependent ensemble average
\begin{equation}
\langle\langle \hat X\rangle\rangle_t=\langle\langle \psi |\hat X|\psi\rangle\rangle=\sum_n p_n(t)\, X_n    
\end{equation}
where $X_n$ is defined as 
\begin{equation}
X_n=\frac{1}{N_n}\sum_\alpha \langle\phi_n^\alpha|\hat X|\phi_n^\alpha\rangle   \end{equation}
and represents the average position inside the bin $\Delta X_n$ which characterizes the single superposed macrostate $|\psi_n\rangle$.
A generic change of the variables $\delta p_n$, with the normalization constraint $\sum_n\delta p_n=0$ would contribute a macroscopic change
of the perceived position
\begin{equation}
\delta\langle\langle \hat X\rangle\rangle_t=\sum_n\delta p_n(X_n-X_k) 
\label{displacement}
\end{equation}
where $X_k$ is arbitrarily chosen. Each of the terms in the sum adds a contribution to the average speed of order
\begin{equation}
\left|\frac{d}{dt}\langle\langle\hat X\rangle\rangle \right| 
\ge\left|\frac{d p_n}{dt}\right| \,\Delta X   
\end{equation}
where $\Delta X$ is the minimal bin size for the position. Even if the contributions are summed algebraically in Eq.(\ref{displacement}),
potentially, each of them adds
a macroscopic observable change of the position 
which might violate causality unless
\begin{equation}
 \left|\frac{d p_n}{dt}\right|\,<\frac{1}{\tau_R}=\frac{v_{MAX}}{\Delta X}<\frac{c}{\Delta X} \,,  
\end{equation}
where $v_{MAX}$ is the maximum effective light cone speed which would be allowed by the actual dynamics of the system\cite{lieb0, lieb1,lieb2}. In the non-relativistic context of any macroscopic real system $v_{MAX}\ll c$ yielding a time scale $\tau_R=\Delta X/v_{MAX}\gg\Delta X/c$.
The {\it exact} dynamics driven by $\hat H$ must comply with that limit
and must enforce a natural maximum for the time derivative of the macroscopic
variables $p_n$. In fact, being related to the macroscopic features of the system,
these variables cannot follow the very fast dynamics of the microscopic
degrees of freedom. Of course, no explicit sign of that time scale is found
in the stochastic equation, Eq.(\ref{stoch1}), where the double limit
$\tau_R, \tau_c\to 0$ is understood. 

According to our estimate of the correlation time in Sec.III, $\tau_c=\hbar/\Delta E$,  we conclude that the It\^o interpretation of the stochastic equation is
correct provided that the limit $\tau_R, \tau_c\to 0$ is taken while
\begin{equation}
\frac{\tau_R}{\tau_c}=\frac{\Delta X/v_{MAX}}{\hbar/\Delta E}\gg 1\, ,
\end{equation}
condition which is equivalent to
\begin{equation}
\Delta X\, \Delta E\gg \hbar\, v_{MAX}  
\label{condition}
\end{equation}
and is easily shown to be fulfilled for any real non-relativistic macroscopic object.

For instance, the quantum state of a macroscopic ball with a mass $M$, oscillating in a harmonic potential with a frequency $\omega$, can be described by a 
coherent state, a minimal Gaussian wave packet, with a Poissonian energy dispersion
$\sigma_E/E=1/\sqrt{n}=\sqrt{\hbar\omega/E}$, where $n$ is the average quantum number and $E\sim n\, \hbar\omega$ is the average energy. The width of the Gaussian packet is given by
the minimal uncertainty $\sigma_X=\sqrt{\hbar/2M\omega}$, yielding a product
\begin{equation}
  \sigma_E\,  \sigma_X=\sqrt{\hbar\omega E}\, \sqrt{\frac{\hbar}{2M\omega}}=\hbar\sqrt{\frac{E}{2M}}=\frac{1}{2}\hbar V_0
\end{equation}
where $E=MV_0^2/2$ when written in terms of the maximum speed amplitude $V_0$.
A superposition of two different coherent states can be perceived as such only if
the difference $\Delta X$ of their average positions is much larger than the width $\sigma_X$, with an energy difference $\Delta E\gg\sigma_E$, yielding
\begin{equation}
\Delta X\, \Delta E\gg  \sigma_E\,  \sigma_X \sim  \hbar V_0\, ,
\end{equation}
thus fulfilling the condition of Eq.(\ref{condition}).

In the relativistic context, the condition of Eq.(\ref{condition}) would be replaced by the stronger constraint 
\begin{equation}
\Delta X\, \Delta E\gg \hbar c\, ,
\label{relcond}
\end{equation} 
which is a sufficient condition (but not necessary) for ordinary non-relativistic macroscopic objects. Eq.(\ref{relcond}) is anyway satisfied by all ordinary macroscopic objects of our real world. Actually, $\hbar c\approx 10^{-26}$~Jm
and Eq.(\ref{relcond}) is satisfied if $\Delta E$ and $\Delta x$ are quite
larger than $10^{-13}$  Joule and meters, respectively. We can hardly reach such accuracy in the measure of any real object.

\section{Discussion}
In summary, we have seen that any superposition of macroscopic quantum states can be described by a stochastic equation which is
a reducing equation, driving the superposition towards a quick collapse, if some conditions are met, say when the bin widths $\Delta E$ and
$\Delta X$ are large enough. That condition, as specified in Eqs.(\ref{condition}),(\ref{relcond}), ensures that the stochastic equation
can be studied as a It\^o SDE, predicting a collapse in a very short time $t\sim \hbar\Delta E/| V_{nm}|^2$ according to Eq.(\ref{collapse}).
In fact, the average matrix element $|V_{nm}|$ is a macroscopic energy, 
arising from the Hamiltonian of the whole macroscopic many body system:
even if defined as off diagonal, it is expected to be of order unity in conventional units, yielding a time $t$ formally shorter than $\tau_c$ at least. On the other hand, the sufficient condition of Eq.(\ref{relcond}) is generally met by what we mean as a macroscopic superposition:
any real superposition of macroscopic quantum states would be perceived as such only if the superposed states have different expectation values, with differences $\Delta A$ large enough to be
observed.  Then, any macroscopic superposition would collapse almost instantly upon formation. 

While the width of the bin extensions $\Delta A$ are not specified in the setup of the model, the reduction
condition $\Delta E\Delta X\gg \hbar c$ would set a limit to the
maximum bin extension which can survive in a superposition, thus
specifying what we actually mean by a forbidden macroscopic superposition. That limit might be used to predict to what extent a superposition can be observed; it might be even checked by experiments\cite{cats}. Actually, a finite reduction time might be even observed if a mesoscopic quantum state could be built with a very large relaxation time, but still extended enough to give rise to collapse. In fact, counterintuitively, the reduction would onset at a smaller size for
larger relaxation times.

It is remarkable that the SDE equation predicts a collapse which
satisfies the Born rule: the quantum state $|\psi\rangle$ ends up
in the single macrostate $|\psi_n\rangle$ with a probability given
by the initial squared amplitude $p_n(0)$. Then, according to the
von Neumann measurement scheme, any microscopic quantum system which
is entangled with the experimental apparatus would lead to
a macroscopic superposition and would collapse almost immediately, with
a probability given by the Born rule. Once proven for a generic macroscopic object, the Born rule holds in any measurement process,
without having to postulate it. From this point of view, it is the experimental apparatus which fails to split, breaking the deterministic
time evolution of the microscopic system which is observed, irrespective of its nature, even when it is highly non-local.
Thus, the stochastic reduction of any macroscopic quantum superposition
would provide a simple dynamical solution of the measurement problem.

It could be argued that, when two superposed states split apart, the off-diagonal matrix
element $V_{nm}$ decreases exponentially, yielding a very large reducing time which would diverge according to Eq.(\ref{collapse}), leading to some  sort of asymptotic freedom. However, since $V_{nm}$ is initially a macroscopic large matrix element of order $E_n$, the reducing time turns out to be really short,  leading to an immediate collapse at the onset of the superposition, before the superposed states can move apart, even if they could move at the speed of light. If the macroscopic states cannot
split at all, no Schr\"odinger cat can be built in any measurement process, yielding the Born rule in all practical cases. 

Nevertheless, despite the potential relevance of the  findings, we must be cautious in drawing conclusions.
We must keep in mind that
the results rely heavily on many general assumptions. Among them, the main ones
are: the complete random nature
of the microstates and their sharp decoupling from the macroscopic variables; the approximation
of the random matrix element $W_{n\, m}$ by a pure white noise; the existence of a relaxation time $\tau_R$ which is much larger than the correlation time $\tau_c$; the existence of an 
off-diagonal matrix element $V_{nm}$ which is still large enough when the macrostates start splitting. While all the assumptions seem reasonable, some experimental tests, in the direction outlined above, would be welcome in order to assess their general validity.

\begin{acknowledgments}
I am grateful to Marcello Baldo for valuable and enlightening discussions and for helpful comments on the manuscript.
\end{acknowledgments}

\end{document}